\documentclass[conference]{IEEEtran}

\IEEEoverridecommandlockouts
\usepackage{cite}
\usepackage{amsmath,amssymb,amsfonts}
\usepackage{algorithmic}
\usepackage{graphicx}
\usepackage{textcomp}
\usepackage{xcolor}
\usepackage{hyperref}
\usepackage[noend]{algorithm2e}
\usepackage{titlesec}
\usepackage{xcolor}

\usepackage{lipsum}% http://ctan.org/pkg/lipsum
\usepackage{graphicx}% http://ctan.org/pkg/graphicx
\def\BibTeX{{\rm B\kern-.05em{\sc i\kern-.025em b}\kern-.08em
    T\kern-.2000em\lower.7ex\hbox{E}\kern-.125emX}}
\begin{document}

\title{Security Orchestration, Automation and Response Engine for Deployment of
Behavioural Honeypots
%{\footnotesize \textsuperscript{*}Note: Sub-titles are not captured in Xplore and
%should not be used}
\thanks{Funded by the Science and Engineering Research Board,
Department of Science and Technology, Government of India}
}

\author{\IEEEauthorblockN{Upendra Bartwal, Subhasis Mukhopadhyay, Rohit Negi, Sandeep Shukla}
\IEEEauthorblockA{\textit{C3i Center, Department of Computer Science and Engineering} \\
\textit{Indian Institute of Technology}\\
Kanpur, India \\
(upenbart, subhasism, rohit, sandeeps)@cse.iitk.ac.in}
}

\maketitle
\thispagestyle{plain}
\pagestyle{plain}
\begin{abstract}
Cyber Security is a critical topic for organizations with IT/OT networks as they are always susceptible to attack, whether insider or outsider. Since the cyber landscape is an ever-evolving scenario, one must keep upgrading its security systems to enhance the security of the infrastructure. Tools like Security Information and Event Management (SIEM), Endpoint Detection and Response (EDR), Threat Intelligence Platform (TIP), Information Technology Service Management (ITSM), along with other defensive techniques like Intrusion Detection System (IDS), Intrusion Protection System (IPS), and many others enhance the cyber security posture of the infrastructure. However, the proposed protection mechanisms have their limitations, they are insufficient to ensure security, and the attacker penetrates the network. Deception technology, along with Honeypots, provides a false sense of vulnerability in the target systems to the attackers. The attacker deceived reveals threat intel about their modus operandi. We have developed a Security Orchestration, Automation, and Response (SOAR) Engine that dynamically deploys custom honeypots inside the internal network infrastructure based on the attacker's behavior. The architecture is robust enough to support multiple VLANs connected to the system and used for orchestration. The presence of botnet traffic and DDOS attacks on the honeypots in the network is detected, along with a malware collection system. After being exposed to live traffic for four days, our engine dynamically orchestrated the honeypots 40 times, detected 7823 attacks, 965 DDOS attack packets, and three malicious samples. While our experiments with static honeypots show an average attacker engagement time of 102 seconds per instance, our SOAR Engine-based dynamic honeypots engage attackers on average 3148 seconds.
\end{abstract} 

\begin{IEEEkeywords}
honeypots, orchestration, deception, soar, ddos
\end{IEEEkeywords}

\section{Introduction}
Honeypots have been in use since the late 90s and are employed to deceive attackers by providing vulnerable and fake systems. It provides an extra layer of deception to the organizations. Honeypots are such a tool that organizations can deploy and weaponize to defend themselves and gather threat intel about the types of attacks. Some challenges that arise when honeypots are deployed:
\begin{enumerate}
\item Studies have shown that the number of attacks on static honeypots decreases with a longer deployment period\cite{Sehgal2020}. Hence, gathering threat intelligence becomes challenging.
\item Detecting attackers inside an organization's internal networks becomes very difficult as the insider attackers are well trained with the systems in place at their organization and can distinguish a honeypot from a real system faster than an outsider attacker. This means that the deception systems need to be robust and intelligent enough to deceive them. 
\item Engaging the attackers becomes an arduous task if honeypots are not deployed dynamically. 
\end{enumerate}
Attacks in IT/OT infrastructure are very organization-specific; hence customized deception is the need of the hour. Though there are different open-source threat intelligence feeds, they are insufficient in providing organization-specific threat intelligence. To protect the organization from insider attacks, we need organization-specific deception technology. 
The best way to implement it is to create a generalized deception solution and then customize it for organizations. We have developed a Security Orchestration, Automation, and Response Engine that dynamically deploys honeypots according to the attacker's behavior in internal networks. We observed that dynamically deployed honeypots attract more attackers. The attacker spends more time exploring them, which results in attacker engagement in the honeypots, hence, alerting the organization's security team and providing them time to safeguard the real systems.

\section{Background and Related Work}

\subsection{Background}
As of 2021, 4.66 billion people worldwide are active Internet users\cite{internet} and 57.7 billion US dollars was spent worldwide on cyber-security\cite{stats}. Organizations and industries provide their services online and have extensive networks to handle their user base, attracting cyber attacks. The organizations are also trying to defend their infrastructure and evolve to prevent and mitigate these cyber attacks. Several commercial and open-source SIEM\cite{siemsolutions}, EDR\cite{edrsolutions}, TIP\cite{tipsolutions} and ITSM\cite{itsmsolutions} systems along with IDS, IPS prevent these attacks. However, these systems have their limitations and fail to detect novel attacks, resulting in attackers penetrating the network\cite{siemfail}\cite{edrfail}\cite{tipfail}\cite{itsmfail}. These defensive systems need reconfiguration every time there is a novel/zero-day attack. For example, the rules in the firewall, IDS, IPS need to be updated, and the machine learning models need retraining. For this, we need continuous monitoring and uninterrupted threat intelligence, which is organization-specific.
SOAR (Security Orchestration, Automation, and Response) systems help minimize these risks as they are a collection of security software solutions and tools for browsing and collecting data from various sources. As per Gartner, SOAR systems enable organizations to collect inputs monitored by the security operations team.
Deception Technology is a critical part of the cyber security infrastructure, as it helps organizations confront attackers. Honeypots are one of the core components of deception technology, and a considerable number of honeypots have been developed over the last 20 years. Honeypots gather threat intel about the attacks and trap the attacker into attacking fake systems, shielding existing systems. Honeypots are the tools to deceive an attacker, and it does that by acting as one of the attacker's targets. Luring an attacker is one side of the coin but engaging it for as long as possible is the other side. Orchestrating the honeypots as per the need ensures the honeypots' credibility and saves resources and time. There has been much work in honeypot development, so there is little room for more novel honeypot development. However, there has been hardly any work on intelligently deploying honeypots in the network. 
Most of the time, honeypots are built and deployed on a system in the network 24 hours a day and seven days a week, which results in wastage of resources if the honeypots remain unused and the identification of honeypots becomes easy. Following are the datasets that we used:
\subsubsection{HTTP Honeypot Host-Based IDS}
We have developed machine-learning models to classify attacks in the HTTP honeypots that are dynamically deployed by the SOAR Engine.
The HTTP Host-Based Intrusion Detection System classifies three HTTP attacks, namely XSS\cite{OWASP}, SQLi\cite{OWASP}, and OSC\cite{OWASP}.
For training, we used ECML/PKDD 2007 Dataset\cite{DS1} and HTTP CSIC Torpeda 2012 Dataset\cite{DS2} and combined them to form a large dataset. The breakdown of the dataset in terms of attack and normal data points:

\begin{table}[htbp]
\vspace{-1em} 
\caption{Dataset Breakdown for HTTP IDS}
\begin{center}
\begin{tabular}{|p{0.3cm}|p{0.7cm}|p{1.1cm}|p{1.1cm}|p{1.1cm}|p{0.9cm}|p{0.6cm}|}\hline
%\begin{tabular}{|c|c|c|c|c|c|c|}
\textbf{S No.} &\textbf{Attack Type} & \textbf{Total Requests} &\textbf{Attack Requests}&\textbf{Benign Requests}&\textbf{Training Set}&\textbf{Test Set}   \\\hline

 1.&XSS &49761& 6573&43188&37320&12441\\\hline
 2.&SQLi &81850&38662&43188&61387&20463 \\\hline
 3.&OSC & 45490&2302&43188&34117&11373\\\hline

\end{tabular}
\label{tab1}
\end{center}
\vspace{-1em} 
\end{table}

%\begin{figure}[h]
%\centering
%\includegraphics[scale=0.8]{datasetbreakdownhttpids.eps}
%\caption{Dataset breakdown for log based HTTP IDS }
%\label{fig:HTTPlog}
%\end{figure}

\subsubsection{Botnet Detection}
We have developed machine-learning models to classify traffic as a botnet or regular traffic as a part of the SOAR Engine. For training, we used the CTU-13 dataset\cite{GARCIA2014100} captured by CTU University, Czech Republic,2011. The purpose of creating the dataset was to capture botnet traffic mixed with regular traffic.

\begin{table}[htbp]
\vspace{-1em} 
\caption{Dataset breakdown for Botnet Detection}
\centering
\begin{tabular}{|p{0.6cm}|p{1.5cm}|p{1.5cm}|p{1.5cm}|p{1.5cm}|}\hline
 \textbf{S No.} &\textbf{Data-point Tag} & \textbf{Data-points(No.)} &\textbf{Training Points} &\textbf{Test Points}   \\\hline
 1&Botnet  &329183 & 247077& 82106\\\hline
 2&Normal & 681495& 510931
& 170564 \\\hline
 3&Total & 1010678& 758008&252670\\\hline
\end{tabular} \newline \newline
\label{tab:honeypot}
\vspace{-3em} 
\end{table}

%\begin{figure}[h]
%\centering
%\includegraphics[scale=0.6]{botnetdatasetbreakdown.eps}
%\caption{Dataset breakdown for Botnet Detection}
%\label{fig:HTTPlog}
%\end{figure}

\subsubsection{DDOS Detection}
We have developed machine-learning models to classify traffic as DDOS or regular traffic as a part of the SOAR Engine.
For training, we used the CIC-IDS 2017 dataset \cite{10.1007/978-3-030-25109-3_9}. The dataset was captured for a week, and each day captured different attacks. The DDOS attack was captured on Wednesday, so for training the model, the PCAP files of Wednesday have been used.

\begin{table}[h]
\vspace{-1em} 
\caption{Dataset breakdown for DDoS Detection}
\centering
\begin{tabular}{|p{0.7cm}|p{1.7cm}|p{1.5cm}|p{1.5cm}|p{1.5cm}|}\hline
 \textbf{S No.} &\textbf{Data Tags} & \textbf{Data-points(No.)}&\textbf{Training Points} &\textbf{Testing Points}\\\hline
 1&DDOS Attack & 656994&349639&307355\\\hline
 2&Normal & 1343006&1045737&290664\\\hline
 3&Total &2000000&1395376&598019 \\\hline

\end{tabular}
\label{tab:honeypot}
\vspace{-1em} 
\end{table}

\subsection{Related Work}

According to our knowledge, there is no prior work reflecting this idea of orchestrating honeypots, some came closest to the concept. This\cite{idsarticle} paper proposed to identify the unused IPs in the network and assign them to the honeypots. They also proposed to use low interaction honeypots, then redirect the incoming packet to the low interaction honeypot, to send them to the high interaction honeypot to get a valid response. This\cite{resularticle} paper proposed the deployment of Honeypots by implementing a graphical user interface to create and destroy a honeypot manually. Another \cite{unknown} paper proposed to deploy honeypot when a decided alarm is triggered; they proposed to use docker for honeypot deployment. Many commercial software solutions \cite{commercialdeception} claim dynamic deception. However, the working of these solutions is not open-source, and no data validating it could be found. The limitations and research gap lies in the fact that there are no methods to dynamically orchestrate a honeypot to ensure that the honeypot's detection is delayed for the maximum time and automate the deployment technique in line with the attacker's interest in the organization.

\section{Problem Statement}
Let us consider an organization with an internal network where several devices are connected to the internet and several air-gapped networks that host several critical applications, protected using an amalgamation of SIEM, EDR, TIP, and ITSM. As discussed in the previous sections, these systems do not guarantee the security of a network and fail to detect novel/zero-day attacks and insider threats\cite{insiderattacksexp}\cite{databreach60}. Air-gapped networks remain vulnerable to attacks as physically detaching the network from the internet does not guarantee its security, as seen in various cases\cite{airgapattack2}.
To mitigate these problems, we need to create a deception defense mechanism that engages the attacker longer and provides uninterrupted organization-specific threat intelligence. We propose a SOAR Engine that:
\begin{enumerate}
    \item Intelligently deploys and deletes honeypots in one or multiple networks, hence attracting more attackers.
    \item Saves CPU time and increases attacker engagement time in the honeypots.
    \item Implement a DDOS attack detection, botnet detection, and malware collection system that monitors the honeypot network for any malicious activity and reports it. 
\end{enumerate}
The SOAR engine adds an extra layer of deception in the internal network along with the existing techniques already in place. The engine also helps save resources and intelligent deception, securing the organization's internal network. With the integration of our novel SOAR engine in deception, organizational networks can have an extra layer of security even if the attacker has infiltrated inside the network undetected.

\section{Proposed Methodology}
We have developed an SOAR Engine that will deploy honeypots according to attacker behavior inside the network. A few IPs inside the network is reserved and used to deploy honeypots. These reserved IPs would not be known to genuine users. The traffic in these IPs is continuously monitored, and whenever there is any incoming traffic to these IPs, it can be pondered that the traffic has malicious intent. The engine would then orchestrate the honeypots as per the traffic and use the response from the attacker to orchestrate the honeypots even further.  Compared with real-life war scenarios, these reserved IPs act as mines for the attacker. This SOAR engine can be considered the last line of defense for an organizational network.
The SOAR Engine involves many components. It is an ensemble of different technologies to develop a system that can be fruitful for organizations to protect their network. This section describes the architecture of the SOAR engine designed and its interaction with the rest of the components.
\subsection{System Design}

\begin{figure}
  \includegraphics[width=8cm,height=4cm]{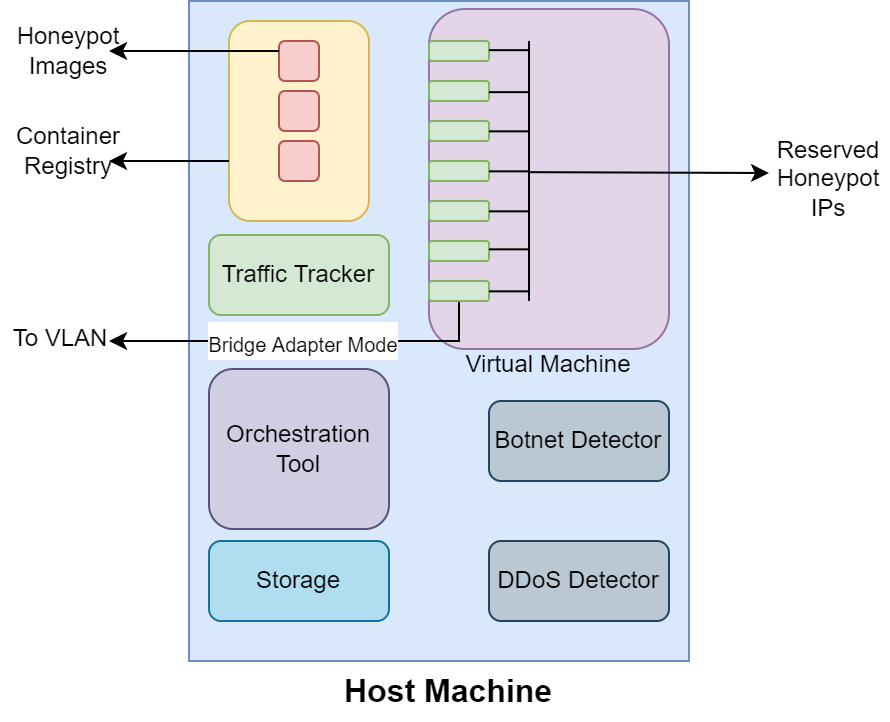}
  \caption{Components of SOAR Engine}
\end{figure}
\begin{enumerate}
    \item  {\textbf{Host Machine}}: Hosts every component.
    \item  {\textbf{Virtual Machine}}: Hosts the honeypots, used to connect multiple VLANs.
    \item  {\textbf{Honeypots}}: Base of the SOAR Engine. Used to trap attackers in the VLAN networks.
    \item  {\textbf{Container Registry}}: Server that hosts the honeypot images.
    \item  {\textbf{Storage}}: Stores logs, backups and files for the SOAR Engine.
    \item  {\textbf{Traffic Tracker}}: Monitors the incoming and outgoing traffic to the reserved IPs in the virtual machine.
    \item  {\textbf{Botnet Detector}}: Detects Botnet traffic to the reserved IPs.
    \item  {\textbf{DDOS Detector}}: Detects DDOS traffic to the reserved IPs.
    \item  {\textbf{Orchestration Tool}}: Orchestrates and automates the other components so that no human involvement is required except for starting the engine. The traffic tracker component captures all the incoming traffic on the host, and based on that; the orchestration tool decides to start the honeypot. Initially, no honeypot is deployed, suggesting no attackers present inside the network.As soon as traffic to the reserved IPs is detected, employing an attacker's presence, the orchestration tool deploys respective honeypots based on the attacker's interaction. As the attacker interacts with the honeypots, it deploys the next honeypots based on their interaction with the previous honeypots to ensure more extended engagement with the honeypots.The orchestration tool uses a hybrid approach (ML and Rule-Based) for decision-making.
    \item  {\textbf{Access Logs}}: Generates logs in every component of the SOAR engine, which is then stored in the storage component.
\end{enumerate}

\begin{figure}
  \includegraphics[width=9.1cm,height=4cm]{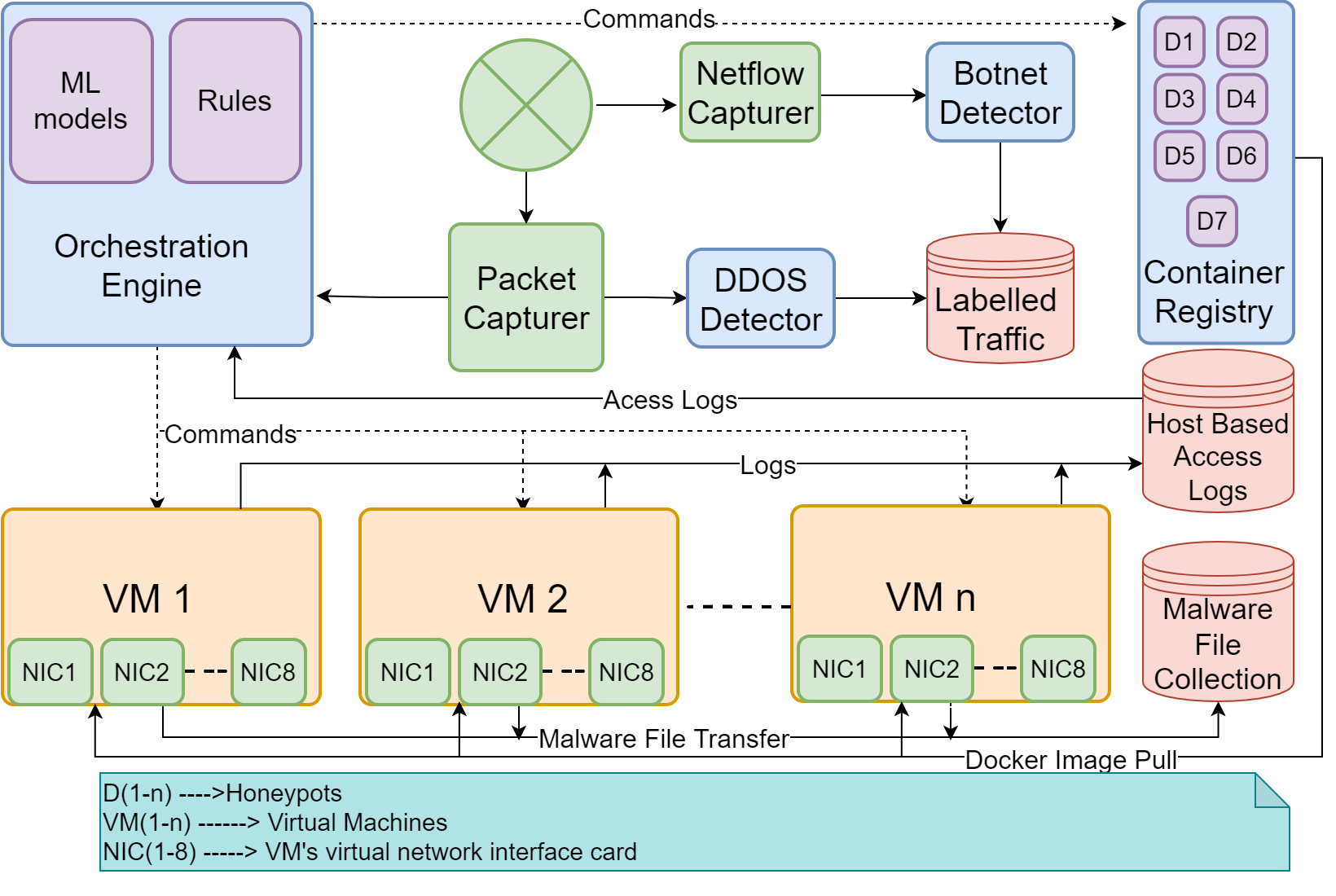}
  \caption{SOAR Engine Architecture}
\end{figure}

\subsection{Implementation}
The strength of a honeypot depends on its design and deployment techniques. As we are inviting a series of deadly attacks on the honeypot systems, hence it should be made sure that the honeypots work reliably. This SOAR engine implements a dynamic honeypot deployment technique intelligently using a set of rules and decisions from machine-learning models.

\subsubsection{Virtual Machine Setup}
A virtual machine is created with Virtualbox with the operating system as Ubuntu Server 18.04 LTS where open-ssh server and docker are installed. By GUI, VirtualBox allows connecting four adapters to the physical ethernet ports of the machine. In order to add the other four adapters, it has been done through the command line using VirtualBox's own set of commands. The VM has eight adapters, seven of them connected to the network in VLAN. One adapter set to host-only transfers logs/files from VM to host.
The virtual machine's VirtualBox image(ova file) exported is used as a \textbf{base image}.
\subsubsection{Network Setup}
The SOAR Engine deployed on a VLAN with seven IPs reserved for deploying honeypots, with the static IPs being on equal intervals from each other distributed throughout the sub-net. For example, a VLAN  having subnet 172.26.233.0, the first IP is reserved at 172.26.233.4, the second IP reserved at 172.26.233.40, the third IP reserved at 172.26.233.85, the fourth IP reserved at 172.26.233.125, the fifth IP reserved at 172.26.233.185, the sixth IP reserved at 172.26.233.220, and the seventh IP reserved at 172.26.233.250. In this way, all the seven IPs are reserved and distributed throughout the sub-net. DHCP does not allocate these IPs to any genuine user/services in the VLAN.
\subsubsection{Working of the SOAR Engine}
On initialization, the SOAR Engine receives input of the number of virtual machines to start, the RAM size of each virtual machine, the number of CPU cores to allocate, whether it should run the previously stopped instances or start fresh, and the host's network interface through which the virtual machine should be connected.

Working of the SOAR Engine in chronological order:
\begin{enumerate}
\item Start the container registry, orchestration tool, DDOS detector, and Botnet Detector on different threads per the inputs received.
\item Transfer all the honeypot images required for the VLAN sub-net from the container registry to the VM.
\item Analyze the incoming packets in the physical ethernet card (NIC) connected to the VLAN while checking some attributes:
	\begin{enumerate}
	\item If the destination IP of the packet belongs to any static IPs assigned to deploy honeypots.
	\item If the destination port is of some application running in the VLAN. For example, if an HTTP server runs on VLAN, there would be an HTTP honeypot in the container registry. So, the SOAR engine will look for port 80 being pinged/accessed in the reserved IPs.
	\item If both the above points are valid; it might be the case that someone is trying to scan the network; it might be that some malicious insider/ attacker is trying to scan the network. The SOAR Engine will trigger to deploy the honeypot for the service that uses the port that came as the packet's destination port. For example, if the packet's destination port is 80, the SOAR Engine will deploy an HTTP honeypot in the next reserved IP. If there is already an HTTP honeypot running in the next reserved IP, the engine will update the latest incoming traffic to the current time, meaning that the honeypot is not idle. Idle honeypots are deleted after being idle for some time.
\end{enumerate}

\item The SOAR Engine will analyze the host-based logs of the deployed honeypots running. For example, if in a VLAN of 172.26.233.0 -- 172.26.233.4, 172.26.233.40, 172.26.233.85, 172.26.233.125, 172.26.233.185, 172.26.233.220, 172.26.233.250 are the reserved IPs, and there is an HTTP honeypot deployed at 172.26.233.4. The orchestration tool will monitor and analyze the logs of the HTTP honeypot and will deploy respective honeypots in the following IP. If there is an SQL Injection attack detected in the HTTP honeypot at 172.26.233.4, then an SQL Injection Honeypot would be deployed at 172.26.233.40. In this way, the attacker would be engaged in attacking the reserved IPs, which act as mines, alerting the organization's security team and provide more time to protect existing infrastructure.
\end{enumerate}

\begin{algorithm}
\SetAlgoLined
\textbf{Let:}\newline
Packet=Incoming packet\newline
IP\_dst= IP on which packet will go \newline
IP\_port= Destination port of incoming packet
IP\_honeypot= $[x_i,x_{i+j},x_{i+2j}, \cdot,x_{i+6j}]$  where j is a fixed interval and x is an IP\newline
Ports=$\{1,2,3\cdot\cdot\cdot 65353\}$\newline
Honeypots= $\{(x,p) \mid$ x is a application and x is exposed on p and $p \in Ports\}$\newline
\textbf{Assumptions:}\newline
All the VMs are started and configured with IP\_honeypot and with host-only IP\newline
\For {each Packet}
{
       \If{$IP\_dst \in IP\_honeypot$}
    {
        \If{$IP\_port \in Ports$ }
    {
        \If{$\exists$ hp such that $hp \in Honeypots$ and $hp.p=IP\_port$}
    {
         IP $\leftarrow$ get the IP for deploying honeypot\newline
        HostIP $\leftarrow$ get the host IP on which IP is assigned\newline
         Deploy Honeypot on IP if not already deployed \newline
         Start supporting modules in a separate thread \newline
    }
    }
    }
    }
\KwResult{Honeypot is deployed}
\caption{Algorithm for honeypot deployment by SOAR Engine}
\end{algorithm}

Among the seven reserved IPs used to deploy the honeypots; there must be a way to deploy the honeypots among the IPs according to the attacker's behavior. During the reconnaissance phase, an attacker scans the network to gather information about the services in the systems deployed/running. While we deploy honeypots, it might be the case that the attacker has already scanned the IP where we have deployed the honeypot, following which the deployment would be of no use.
We have developed an algorithm that will deploy the honeypots per the attacker's behavior. There is no redundancy, and honeypots are not deployed in the reserved IPs already scanned. \textit{Algorithm 2} shows the IP allocation algorithm we have developed. \newline
\begin{algorithm}
\SetAlgoLined
\textbf{Let:}\\ 
IP\_dst= IP on which packet came \\
IP\_honeypot= $[x_i,x_{i+j},x_{i+2j}, \cdot,x_{i+6j}]$  where j is a fixed interval and x is an IP and i a constant showing first IP\\
IP\_assigned= $\{x \mid x \in $IP\_honeypot and no honeypot is deployed in $x\}$\\
IP\_no= Number of IP to return  \\
\eIf{IP\_dst is first IP of IP\_honeypot}{
IP\_list $\leftarrow$ IP from IP\_honeypot except IP\_dst\\
           
IP\_list $\leftarrow$ IP\_list-IP\_assigned\\
return first IP\_no number of IP from IP\_list 
}
{
IP\_list $\leftarrow$ IP from IP\_honeypot except IP\_dst\\
           
IP\_list $\leftarrow$ IP\_list-IP\_assigned\\
return last IP\_no number of IP from IP\_list \newline
}
\KwResult{Reserved IPs that are free currently }
 
\caption{Algorithm for IP selection to deploy honeypots in reserved IPs}
\end{algorithm}

The honeypots deployed are continuously checked for new files added. The new files are sent to the storage component repository, from where it is sent for further analysis. The honeypots are also checked for being active. The metadata of deployed honeypot with the timestamp of the last incoming packet is checked with the time the honeypot was deployed. If the honeypot has no activity for the last 15 minutes, a backup of the honeypot image is stored, and then it is deleted.
\begin{figure}
  \includegraphics[width=9cm,height=4cm]{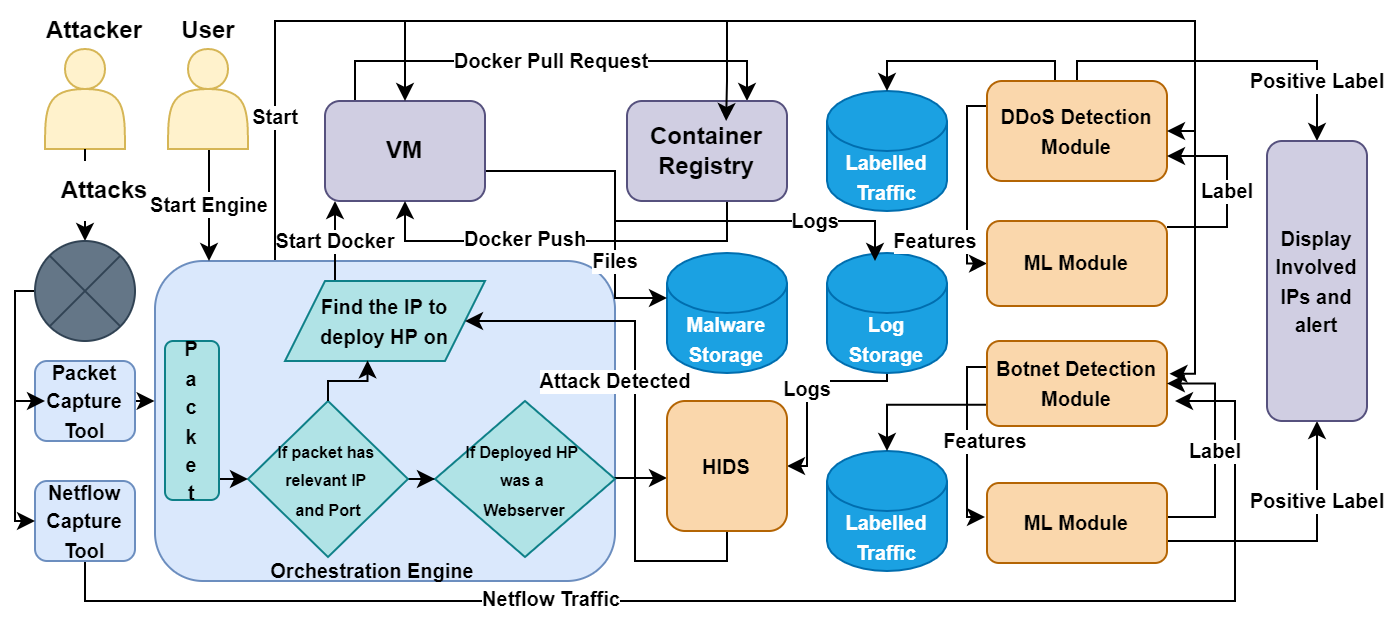}
  \caption{Execution Chart of the SOAR Engine}
\end{figure}

Since our primary goal is to deploy honeypots intelligently, we used modified versions of some open-source honeypots. All of them have been dockerized to ensure they are lightweight and could restart quickly in case of any failure. The honeypots we used are:

\begin{enumerate}
\item Web Server | \textbf{apache/httpd} | Apache Server Docker used as high interaction HTTP honeypot.
\item Application Server | \textbf{Tomcat} | Tomcat Server docker for high interaction HTTP application server honeypot.
\item Database Server | \textbf{MySQL} | MySQL Server version 5.7 docker for high interaction database server honeypot.
\item SSH | \textbf{Cowrie} | Medium to high interaction open-source honeypot
\item SMTP |  \textbf{Mailoney} | Open-Source SMTP Honeypot written in python
\item Modbus | \textbf{Custom Modbus Server} | Modbus docker image used for high interaction modbus honeypot.
 
\end{enumerate}
\subsubsection{HTTP IDS, Botnet, and DDoS Detection}
In a real-world scenario, several applications interact to make a proper solution. The HTTP honeypot deployed is designed to display a page that takes input from the user and stores it in a database. Hence, the Tomcat Server, HTTP Server, and the database server are used. There are also three other HTTP honeypots with specific vulnerabilities in the container registry: \textbf{SQL Injection} (effects database server)\cite{OWASP}, \textbf{Cross-Site Scripting} (effects application server)\cite{OWASP}, and \textbf{OS Command Injection} (effects application server)\cite{OWASP}. \newline
We modified the IDS developed by Bhagwani Et Al.\cite{10.1007/978-3-030-35869-3_10} to predict SQLi, XSS, and OSC attacks on HTTP servers. The SOAR Engine uses these machine learning models to detect attacks in the logs of the initial HTTP honeypot deployed and then deploy HTTP honeypots in other IPs with specific vulnerabilities like SQLi, XSS, and OSC. Different Machine Learning models are implemented and cross validated, and based on the result of the cross validation set the models that gave the best results are selected. 

\paragraph{\textbf{HTTP IDS}}
\textbf{\textit{ Features:}}
All the attributes in the features are frequency-based. Hence we deleted some features since the sum of the frequency was zero. Table IV shows the features that we deleted:

\begin{table}[h]
\vspace{-1em} % <--- this is new
\caption{Excluded features from the referenced attributes}
\centering
\begin{tabular}{|p{0.8cm}|p{0.7cm}|p{6cm}|}\hline
\textbf{S No} &\textbf{Attack} & \textbf{Features}  \\\hline
 1&XSS & ' !','$\wedge$','<>','[]','createelement',
             'search', 'eval()' ,'string.fromcharcode'\\\hline
    2&SQLi & '-',"/**/","'", ';', '\#', '[', ']', '(', ')', '$\wedge$', '|',  '<>', '<=', '>=', '\&\&', '||', ':',' !=','()', \\\hline       
 3&OSC &  '..\textbackslash\textbackslash',  '\textbackslash\textbackslash.', '\textbackslash\textbackslash/',':/','etc/passwd', '`'  \\\hline
\end{tabular}

\label{tab:attkattrib}
\vspace{-1em} % <--- this is new
\end{table}

\textbf{\textit{Detection Models:}} There was a case of data imbalance for XSS and OSC. The class\_weight attribute of each algorithm was used, which deals with the imbalance. In some cases, the accuracy of our models was better than the referenced paper's.
In table V,VI,VII row 1,3,5 shows the accuracy computed by our model, whereas rows 2,4,6 show accuracy computed by the referenced paper.

\begin{table}[htbp]\footnotesize
\vspace{-1em} % <--- this is new
\caption{ML models and accuracy for XSS (in \%)}
\centering
\begin{tabular}{|p{0.7cm}|p{1.7cm}|p{1.2cm}|p{1.2cm}|p{0.8cm}|p{1.0cm}|p{0.5cm}|p{0.5cm}|p{0.5cm}|p{0.5cm}|p{1.2cm}|}\hline
 \textbf{S No.} &\textbf{Classifier} & \textbf{Accuracy} & \textbf{Precision} & \textbf{Recall}  & \textbf{F-Score} \\\hline
 1&Decision Tree & 98.81 &98.83 &98.81 &98.79 \\\hline
  \textit{2}&\textit{Decision Tree} & \textit{99.27} & \textit{98.4}& \textit{97.55}& \textit{97.98}\\\hline
 3&SVM &98.191&98.2&98.19&98.14 \\\hline
  \textit{4}&\textit{SVM} &\textit{98.66}&\textit{98.32}&\textit{98.88}&\textit{98.6} \\\hline
 5&LR & 98.47&98.49&98.47&98.43\\\hline
  \textit{6}&\textit{LR} &\textit{98.04}&\textit{98.4}&\textit{97.55}&\textit{97.98}\\\hline
\end{tabular}
\label{tab:forpol}
\vspace{-1em} % <--- this is new
\end{table}

\begin{table}[htbp]\footnotesize
\vspace{-1em} % <--- this is new
\caption{ML models and accuracy for SQLi (in \%)}
\centering
\begin{tabular}{|p{0.7cm}|p{1.7cm}|p{1.2cm}|p{1.2cm}|p{0.8cm}|p{1.0cm}|p{0.5cm}|p{0.5cm}|p{0.5cm}|p{0.5cm}|p{1.2cm}|}\hline
 \textbf{S No} &\textbf{Classifier} & \textbf{Accuracy} & \textbf{Precision} & \textbf{Recall}  & \textbf{F-Score} \\\hline
 1&Decision Tree & 99.06 &99.08 &99.06 &99.06 \\\hline
 \textit{2}&\textit{Decision Tree} &\textit{ 96.78} &\textit{ 95.83}&\textit{ 97.95}&\textit{ 96.88}\\\hline
 3&SVM &98.12&98.14&98.12&98.12 \\\hline
 \textit{4}&\textit{SVM} &\textit{95.57}&\textit{95}&\textit{96.44}&\textit{95.71} \\\hline
 5&LR & 98.41&98.42&98.41&98.41\\\hline
 \textit{6}&\textit{LR} &\textit{95.7}&\textit{95.16}&\textit{96.53}&\textit{95.84}\\\hline
\end{tabular}
\label{tab:forpol}
\vspace{-1em} % <--- this is new
\end{table}

\begin{table}[htbp]\footnotesize
\caption{ML models and accuracy for OSC (in \%)}
\centering
\begin{tabular}{|p{0.7cm}|p{1.7cm}|p{1.2cm}|p{1.2cm}|p{0.8cm}|p{1.0cm}|p{0.5cm}|p{0.5cm}|p{0.5cm}|p{0.5cm}|p{1.2cm}|}\hline
 \textbf{S No} &\textbf{Classifier} & \textbf{Accuracy} & \textbf{Precision} & \textbf{Recall}  & \textbf{F-Score} \\\hline
 1&Decision Tree & 98.57 & 98.58&98.57 &98.46 \\\hline
 \textit{2}&\textit{Decision Tree} &\textit{ 97.29} &\textit{ 98.61}&\textit{ 94.32}&\textit{ 96.42}\\\hline
 3&SVM &98.02&97.99&98.02&97.81 \\\hline
 \textit{4}&\textit{SVM} &\textit{97.95}&\textit{99.4}&\textit{96.36}&\textit{97.85} \\\hline
 5&LR & 98.44&98.46&98.44&98.31\\\hline
 \textit{6}&\textit{LR} &\textit{97.85}&\textit{99.01}&\textit{96.53}&\textit{97.75}\\\hline
\end{tabular}
\label{tab:forpol}
\vspace{-1em} % <--- this is new
\end{table}

\paragraph{\textbf{Botnet Detection}}
The botnet detection module is responsible for getting net flows in the network and classify as botnet flow or standard flow. Argus server is started, which collects the net flows for one minute, and the total flow during this interval is classified as a botnet or normal flow. If botnet flow is detected, the IP address of the source and destination IP is extracted and notified.

\textit{\textbf{Netflow Format:}}
The argus server creates this net flow file after being fed with the attributes required to detect botnet traffic. A flow is defined by the IP pair and port number pair. 
% The net flow file contains information about the flow. The attributes we need to capture in order are:
% StartTime,Dur,Proto,SrcAddr,Sport,Dir,DstAddr,Dport,\\
% State,sTos,dTos,TotPkts,TotBytes,SrcBytes,Label\\

% Following is an example how a record in a net flow file looks like - \newline
% $2011/08/10 09:46:59.607825,1.026539,tcp,94.44.127.113,\\1577,  ->,147.32.84.59,6881,S\_RA,0,0,4,276,156,flow=Background-Established-cmpgw-CV$
\textit{\textbf{Features:}}
The features used for botnet traffic classification inspired by this reference\cite{git1}, we modified to fit our requirement of collecting the traffic flow for a minute. The features and description of the features used are:
Duration for the transfer of packet, protocol used, source port, destination port, source IP, destination IP, total bytes transferred, state of the netflow entry and total packets transferred.
If duration of packet or byte transfer is greater than 1 minute, then it will be taken as total packet/byte per minute.
%\begin{table}[h]
%\caption{Features used in botnet detection\cite{git1}}
%\centering
%\begin{tabular}{|p{0.4cm}|p{1.2cm}|p{5.0cm}|}\hline
% \textbf{S No} &\textbf{Features} & \textbf{Description}  \\\hline
% 1&Duration & Duration for the transfer of packet happen\\\hline
 %2&Protocol & Protocol used \\\hline
 %3&S Port & Source port \\\hline
% 4&D port& Destination port\\\hline
% 5&S IP & source IP\\\hline
% 6&D IP & Destination IP\\\hline
% 7&Totb & Total byte transferred , if duration greater then 1 min, then total byte per min.\\\hline
% 8&State & State of the netflow entry\\\hline
% 9&Totp & Total packet transferred, if duration greater then 1 min , then total packet per min.\\\hline
%\end{tabular}
%\label{tab:honeypot}
%\end{table}

\textit{\textbf{Detection Models:}} Our Botnet Detection technique provides better accuracy than existing methods\cite{9299061}. The highest accuracy in the referenced method is 99.89\%, whereas the highest accuracy achieved by our method is 99.95\%.

\begin{table}[h]
\vspace{-1em} % <--- this is new
\caption{ML models  and Accuracy of Botnet detection model (in \%)}
\centering
\begin{tabular}{|p{0.7cm}|p{1.7cm}|p{1.2cm}|p{1.2cm}|p{0.8cm}|p{1.0cm}|p{0.5cm}|p{0.5cm}|p{0.5cm}|p{0.5cm}|p{1.2cm}|}\hline
 \textbf{S No} &\textbf{Classifier} & \textbf{Accuracy} & \textbf{Precision} & \textbf{Recall}  & \textbf{F-Score} \\\hline
 1&Decision Tree & 99.95 &99.95 &99.95 &99.95 \\\hline
 2&SVM &97.55&97.68&97.55&97.57 \\\hline
       \end{tabular}
\label{tab:exattrib}
\vspace{-1em} % <--- this is new
\end{table}

\paragraph{\textbf{DDoS Detection}}
The DDOS detection module classifies each packet into a DDOS packet or a regular packet. The source and destination IP are extracted and notified if a DDOS packet is detected. 

\textit{\textbf{Features:}}
Features inspired by this reference\cite{git2}, we have added features like protocol used, source port, destination port, and packet length. The reason behind adding the features are:
\begin{enumerate}
\item Protocol: Protocols like UDP, TCP ICMP are used to attack services.

\item Port: A port signifies which service is getting attacked.

\item Packet Length: There might be cases where the packets may or may not have any data.
\end{enumerate}

\textbf{Features used are:}
Ethernet Source Occurrence-look back 100, Ethernet Destination Occurrence-look back 100, IP Source Occurrence-look back 100, IP Destination Occurrence-look back 100, Ethernet Source Occurrence-look back 1000, Ethernet Destination Occurrence-look back 1000, IP Source Occurrence-look back 1000, IP Destination Occurrence-look back 1000, Timestamp 1: Current packet time - Previous packet time, Timestamp 2: Current packet time - $10^{th}$ packet's time, Timestamp 3: Current packet time - $100^{th}$ packet's time, Timestamp 4: Current packet time - $1000^{th}$ packet's time, Protocol used i.e UDP,TCP or ICMP, Source Port: Source port in packet, Destination port in packet, Length of packet

Occurrence look back 'n' means occurrence of IP address  in last 'n' packets.

\textit{\textbf{Detection Models:}} Our DDOS Detection technique provides better accuracy than existing methods\cite{proceedings2020063051}. The highest accuracy in the referenced method is 97.86\%, whereas the highest accuracy achieved by our method is 99.94\%.

\begin{table}\small
\caption{ML models  and Accuracy of DDOS detection model (in \%)}
\centering
\begin{tabular}{|p{0.8cm}|p{1.8cm}|p{1.2cm}|p{1.2cm}|p{0.8cm}|p{1.2cm}|p{0.5cm}|p{0.5cm}|p{0.5cm}|p{0.5cm}|p{1.2cm}|}\hline
 \textbf{S No.} &\textbf{Classifier} & \textbf{Accuracy} & \textbf{Precision} & \textbf{Recall}  & \textbf{F-Score} \\\hline
 1&Decision Tree & 99.92 &99.94 &99.85 &99.93 \\\hline
 2&SVM &99.92&99.92&99.92&99.92 \\\hline
 3&LR & 99.91&99.91&99.91&99.91\\\hline
\end{tabular}
\label{tab:exattrib}
\vspace{-2em} % <--- this is new
\end{table}

\section{Validation}
\subsection{Experimental Setup}
The SOAR Engine developed was tested in a real network to validate that the system works and is not only in theory. An internal network environment configured to ensure real-life setup and attack scenarios. We used a 32 core processor with a 64 GB RAM machine hosted on DigitalOcean. Since the SOAR engine mainly focuses on deploying honeypots in internal networks, we took advantage of an ongoing Capture The Flag event hosted by the TalentSprint-IIT Kanpur advance certification program. The event involved 66 players playing the whole event. Each player was given a machine inside the network with a unique username and password. The machines with unique IPs given to the players were docker images deployed on the host machine.
In between IPs given to players, some IPs were reserved to be used for honeypot deployment by the SOAR Engine. Also, there were different HTTP, FTP, Modbus, SMTP servers running in some IPs in the network.
Each player can be considered an attacker who has compromised an internal machine and tried to attack the network. The players were given a problem statement where they had to find some flags hidden in files inside some machines in the network. They had to act as attackers trying to attack the whole network and find critical data to solve it. 
This setup ran for four days, and the generated results proved the practical implementation of the SOAR Engine, which was pretty efficient.

\subsection{Results}
\subsubsection{Number of times honeypots deployed:}
One of the essential features of the SOAR Engine is to deploy a honeypot when required and to delete it after it is idle for a long time. This technique helps save resources in the machine where the SOAR Engine is running. The statistics of the Honeypots deployed by the engine are:
\begin{enumerate}
\item Apache -- Deployed and deleted thirteen times
\item SSH -- Deployed and deleted ten times
\item SMTP -- Deployed and deleted ten times
\item Modbus --  Deployed and deleted once
\item XSS, OSC -- Deployed and deleted three times
\item SQLi -- Deployed and deleted three times
\end{enumerate}
The Modbus Honeypot was deployed once in four days. Deploying it statically would have led to wasting of resources.

\begin{figure}[h]
\centering
\includegraphics[scale=0.35]{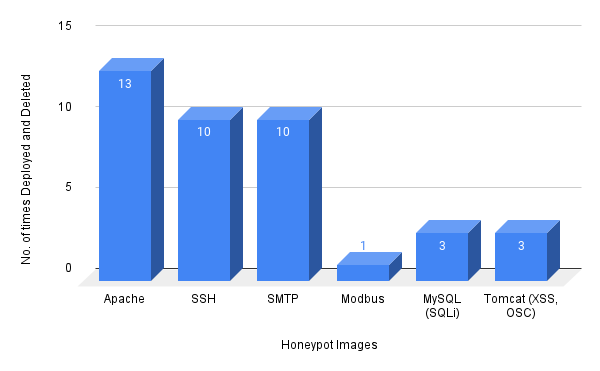}
\caption{Number of times Honeypots were deployed }
\label{fig:sysconf}
\end{figure}

\begin{figure}[h]
\centering
\includegraphics[scale=0.35]{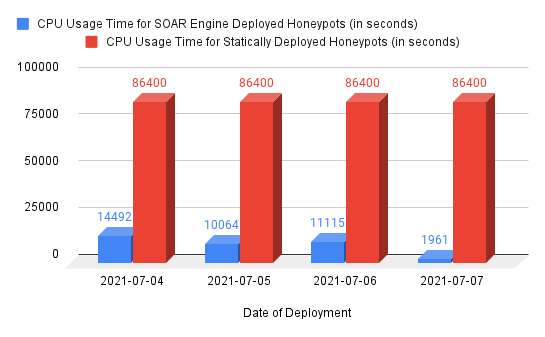}
\caption{CPU Usage Time of SOAR Engine vs Statically Deployed Honeypots}
\label{fig:sysconf}
\end{figure}

\subsubsection{Number of Attacks on the Honeypots:}
In the Capture The Flag event, one of the problem statements required targeting a web server hosting a website. Hence, attacks in the web-based honeypot are much more. However, attacks were also detected in the SSH, MODBUS, and SMTP honeypots. Another problem statement in the CTF required every player to visit a machine assigned to them for solving problems and finding flags. The players also had to act as attackers and visit other machines in the network to find flags and other important intelligence. During this exercise, in addition to finding flags, if they visited another player's machine and attacked it, the attacker was given a positive point, and the victim was given a negative point. If they visited the honeypots and interacted, they were given a negative point. Since there were 66 players and each had a web server running in their assigned machine, every web server can be considered a static honeypot. In the SOAR Engine, the HTTP protocol(web-server honeypot) had two different honeypot images, and on each instance of attack, alternate honeypot images were deployed and deleted(when idle) automatically. It was seen that 66 static web-server honeypots collected 63108 attacks, whereas two web-server honeypot images deployed dynamically collected 7555 attacks. Hence, when the honeypots are deployed statically, each collected 956 attacks, but when deployed dynamically, each collected around 3777 attacks, with the additional benefit of resource-saving when there are no attacks.

% \begin{table}[h]
% \caption{Number of attacks on each Honeypots}
% \centering
% \begin{tabular}{|p{0.2cm}|p{3cm}|p{3cm}|}\hline
%  \textbf{S No} &\textbf{Honeypot} & \textbf{Number of Attacks}  \\\hline
%  1&HTTP & 7555  \\\hline
% 2&SSH & 187  \\\hline
%  3&SMTP & 80  \\\hline
%  4&MODBUS & 1  \\\hline
%  5&Statically Deployed Honeypots & 63108   \\\hline
 
% \end{tabular}
% \label{tab:honeypot}
% \end{table}

\begin{figure}[h]
\centering
\includegraphics[scale=0.3]{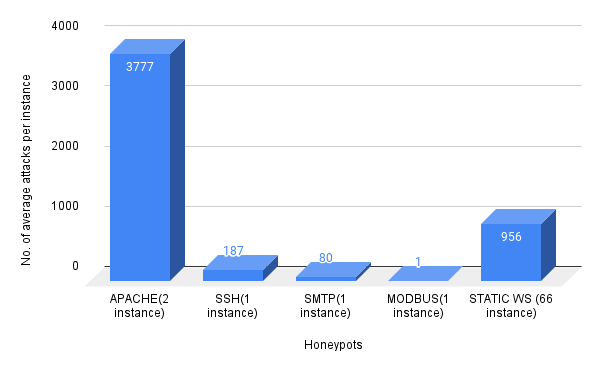}
\caption{Number of Attacks per honeypot image}
\label{fig:sysconf}
\end{figure}

\begin{figure}[h]
\centering
\includegraphics[scale=0.3]{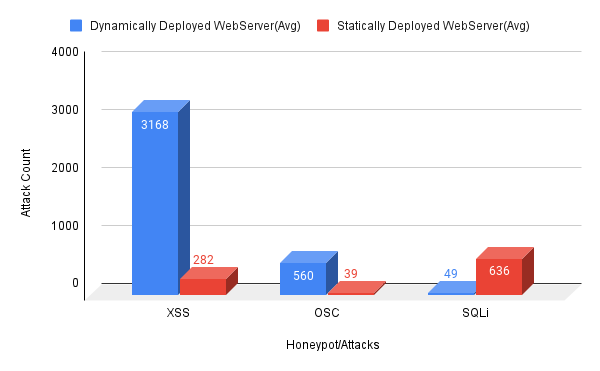}
\caption{Number of Attacks of each type per honeypot image}
\label{fig:sysconf}
\end{figure}

\subsubsection{Comparison of attacks in statically deployed honeypots v/s SOAR Engine deployed honeypots:}

We also deployed static honeypots (of the same protocols used in SOAR Engine) in DigitalOcean to compare the engagement time of the attackers with our SOAR Engine. The honeypots were flagged as honeypots within 30 minutes of deployment, along with the attacker interaction and engagement declining within time. Table X shows the comparison of attacker engagement time of our SOAR Engine with that of honeypots deployed statically.

The results show that statically deployed honeypots attracted fewer attackers and used too many resources than the SOAR Engine deployed honeypots. Hence, it can be established that honeypots deployed by our SOAR engine increase the attacker engagement time and save resources, thereby increasing the efficiency of honeypot deployment. Since these honeypots are deployed only in an attacker's presence, they can be established as behavioural honeypots.

% \begin{table}[h]
% \caption{Statically Deployed Honeypots Vs SOAR Engine Deployed Honeypots}
% \centering
% \begin{tabular}{|p{0.2cm}|p{1.5cm}|p{0.5cm}|p{0.5cm}|p{0.5cm}|p{0.8cm}|}\hline
%  \textbf{S No} &\textbf{Honeypot\newline/Attack} & \textbf{XSS} &\textbf{SQLi} &\textbf{OSC} &\textbf{Normal} \\\hline
%  1&Dynamically Deployed Honeypots & 6336&98&1121&231502  \\\hline
% 2&Statically Deployed Honeypots & 18599  &41971&2538&219190\\\hline
% \end{tabular}
% \label{tab:honeypot}
% \end{table}
\begin{table}[htbp]
\vspace{-2em} % <--- this is new
\centering
\caption{Top 10 Engagement Times in SOAR Engine vs Statically Deployed Honeypots}
\label{tab:my-table}
\begin{tabular}{|p{1.45cm}|p{1.5cm}|p{1.75cm}|p{1.5cm}|}\hline
\multicolumn{2}{|c|}{\textbf{SOAR Engine Deployed Honeypots}} & \multicolumn{2}{|c|}{\textbf{Statically Deployed Honeypots}}   \\ \hline
\textbf{Attacker IP}  & \textbf{Time} & \textbf{Attacker IP} & \textbf{Time} \\ \hline
172.16.238.5  & \textcolor{red}{5977 sec} & 20.55.53.144    & 321 sec \\ \hline
172.16.238.24 & \textcolor{red}{4390 sec} & 104.155.181.214 & 263 sec \\ \hline
172.16.238.58 & \textcolor{red}{4348 sec} & 162.158.167.231 & 165 sec \\ \hline
172.16.238.8  & \textcolor{red}{3964 sec} & 106.208.155.125 & 75 sec  \\ \hline
172.16.238.5  & \textcolor{red}{3413 sec} & 106.208.154.240 & 55 sec  \\ \hline
172.16.238.30 & \textcolor{red}{2069 sec} & 142.93.157.218 & 50 sec  \\ \hline
172.16.238.24 & \textcolor{red}{1987 sec} & 162.158.165.53 & 35 sec  \\ \hline
172.16.238.40 & \textcolor{red}{1961 sec} & 52.136.124.138 & 29 sec  \\ \hline
172.16.238.5  & \textcolor{red}{1871 sec} & 45.146.164.110 & 17 sec  \\ \hline
172.16.238.5  & \textcolor{red}{1509 sec} & 45.146.164.110 & 17 sec  \\ \hline
\end{tabular}
\vspace{-2.5em} % <--- this is new
\end{table}

\subsubsection{DDOS, Botnet Attacks, and Malware Collection:}
The SOAR Engine captured a total of 965 DDOS packets in the network. There was no Botnet positive flow detected in the network during the experiment.
Three malicious samples were also captured by the honeypots deployed by the SOAR Engine. The samples were shell files that tried to:
\begin{enumerate}
    \item Delete the filesystem by using the command rm -rf
    \item Tried to view the /etc/passwd file
    \item Tried to view SSH logs at /var/log/auth.log
\end{enumerate}

\subsubsection{Ready-to-response time of the SOAR Engine}
\paragraph{Selection of IP and Services w.r.t Lateral Movement:}
Our IP selection algorithm selects the IP based on the first IP that the attacker tries to recon. It might be the case that while the honeypot is being deployed, the attacker's scan passes through the IP where the honeypot is to be deployed. The SOAR Engine's ready-to-response time does not allow this. If the \textbf{engine} is deployed in a machine that has \textbf{two cores} and \textbf{4 GB of RAM}, it takes around \textbf{6 seconds} to deploy a honeypot after detecting recon activity on the first IP. 
A simple network scan where the first thousand ports of each system are scanned takes around 1.5 seconds per system. Since the IPs are allocated at a fixed distance, in our experiment, 20 IPs apart, it gives around 30 seconds before the attacker's recon activity reaches the honeypot.
As the system's resources where the SOAR Engine is deployed increase, the engine's performance also increases simultaneously.

\section{Conclusion}
Attacks on organizations are increasing daily, and the type of attacks are organization-specific. Deception Technology has been used for many years to collect threat intelligence about the types of attacks and to deceive the attackers from the original target. Static honeypots do not get engaged enough to know the attacker's capability and modus operandi. Hence, we proposed a Security Orchestration, Automation, and Response Engine to dynamically deploy honeypots as per the attacker's behavior, save resources, and increase the attacker's engagement in the honeypots. It also implemented a botnet and DDOS detection tool for the honeypot network and a malware storage system. The orchestration uses both rule-based and machine-learning techniques for the task. After deploying the whole system in a live environment, we saw that the SOAR Engine deployed honeypots provides a far better attacker engagement inside the honeypots and save around 89\% of the CPU time in the machine deployed.  Three malicious samples were collected by the honeypots and DDOS traffic was also detected. These data and experiments validate our claims that the SOAR Engine performs better than existing systems and can be used by organizations to protect their internal networks.

\section{Future Work}
Honeypots are one of the best information-gathering and threat intel-gathering systems. This paper deals with the idea that if honeypots are deployed intelligently, it can help protect the network infrastructure from being attacked and alert the security team to take action early. The IP selection algorithm can be improved based on the incoming network and port scans. Also, in cases where insider attackers are trying to compromise the critical data in the network have less chance of being detected. The honeypot bank must be increased, and high-interaction honeypots of various honeypots can be developed. Also, attackers may perform lateral movement where they can hop from one IP to the other, so there has to be some way developed to provide a much more guided way to ensure better engagement of the attacker.

%\nocite{*}
\bibliographystyle{plain}
\bibliography{ref}

\begin{thebibliography}{10}

\bibitem{DS2}
Ecml/pkdd 2007 dataset : \url{https://bit.ly/3Evg8Rw}.

\bibitem{DS1}
Http-csic-torpeda-2012 \url{https://bit.ly/3FwB2RB}.

\bibitem{9299061}
Mustafa Alshamkhany, Wisam Alshamkhany, Mohamed Mansour, Mueez Khan, Salam
  Dhou, and Fadi Aloul.
\newblock Botnet attack detection using machine learning.
\newblock In {\em 2020 14th International Conference on Innovations in
  Information Technology (IIT)}, pages 203--208, 2020.

\bibitem{idsarticle}
Hassan Artail, Haidar Safa, Malek Sraj, Iyad Kuwatly, and Zaid Al-Masri.
\newblock A hybrid honeypot framework for improving intrusion detection systems
  in protecting organizational networks.
\newblock {\em Computers and Security}, 25:274--288, 06 2006.

\bibitem{resularticle}
Muhammet Baykara and Resul Das.
\newblock A novel honeypot based security approach for real-time intrusion
  detection and prevention systems.
\newblock {\em Journal of Information Security and Applications}, 41:103--116,
  08 2018.

\bibitem{10.1007/978-3-030-35869-3_10}
Harsh Bhagwani, Rohit Negi, Aneet~Kumar Dutta, Anand Handa, Nitesh Kumar, and
  Sandeep~Kumar Shukla.
\newblock Automated classification of web-application attacks for intrusion
  detection.
\newblock In Shivam Bhasin, Avi Mendelson, and Mridul Nandi, editors, {\em
  Security, Privacy, and Applied Cryptography Engineering}, pages 123--141,
  Cham, 2019. Springer International Publishing.

\bibitem{edrsolutions}
Cynet.
\newblock Edr solutions \url{https://bit.ly/3HgaULa}.

\bibitem{insiderattacksexp}
ekransystem.
\newblock Insider attack \url{https://bit.ly/3z4exAJ}.

\bibitem{commercialdeception}
em360tech.
\newblock Commercial deception \url{https://bit.ly/3FNegow }.

\bibitem{tipsolutions}
esecurityplanet.
\newblock Tip solutions \url{https://bit.ly/3yYzE7K}.

\bibitem{GARCIA2014100}
S.~García, M.~Grill, J.~Stiborek, and A.~Zunino.
\newblock An empirical comparison of botnet detection methods.
\newblock {\em Computers and Security}, 45:100--123, 2014.

\bibitem{databreach60}
Nestor Gilbert.
\newblock Insider threat statistics : \url{https://bit.ly/3qL6vJF}.

\bibitem{itsmsolutions}
guru99.
\newblock Itsm solutions \url{https://bit.ly/32rOtUs}.

\bibitem{siemfail}
Imgainenext.
\newblock Siem failure \url{https://bit.ly/3JjOLgO}.

\bibitem{itsmfail}
Manageengine.
\newblock Itsm fail \url{https://bit.ly/33Q6axc}.

\bibitem{git1}
Nagabhushan.
\newblock ml-based-botnet-detection : \url{https://bit.ly/3yYz6Pe}.

\bibitem{edrfail}
Onsystemlogic.
\newblock Edr failure \url{https://bit.ly/3yZyxEK}.

\bibitem{OWASP}
owasp.org.
\newblock \url{https://bit.ly/33Q52cW}.

\bibitem{airgapattack2}
PortSwigger.
\newblock Air gap attack 2 \url{https://bit.ly/3HkSadz}.

\bibitem{git2}
James Quintero.
\newblock Ddos-attack-detection : \url{https://bit.ly/3Hg7vfa}.

\bibitem{unknown}
Daniel Reti and Norman Becker.
\newblock Escape the fake: Introducing simulated container-escapes for
  honeypots, 04 2021.

\bibitem{proceedings2020063051}
Swathi Sambangi and Lakshmeeswari Gondi.
\newblock A machine learning approach for ddos (distributed denial of service)
  attack detection using multiple linear regression.
\newblock {\em Proceedings}, 63(1), 2020.

\bibitem{Sehgal2020}
Rohit Sehgal, Nishit Majithia, Shubham Singh, Sanjay Sharma, Subhasis
  Mukhopadhyay, Anand Handa, and Sandeep~Kumar Shukla.
\newblock {\em Honeypot Deployment Experience at IIT Kanpur}, pages 49--63.
\newblock Springer Singapore, Singapore, 2020.

\bibitem{10.1007/978-3-030-25109-3_9}
Iman Sharafaldin, Arash Habibi~Lashkari, and Ali~A. Ghorbani.
\newblock A detailed analysis of the cicids2017 data set.
\newblock In Paolo Mori, Steven Furnell, and Olivier Camp, editors, {\em
  Information Systems Security and Privacy}, pages 172--188, Cham, 2019.
  Springer International Publishing.

\bibitem{stats}
Statista.
\newblock Cyber security spending worldwide: \url{https://bit.ly/3ex1Vso}.

\bibitem{internet}
Statista.
\newblock Internet active user stats: \url{https://bit.ly/33OtX0v}.

\bibitem{siemsolutions}
DNS Stuff.
\newblock Siem solutions \url{https://bit.ly/3za1FcE}.

\bibitem{tipfail}
Techtarget.
\newblock Tip fail \url{https://bit.ly/3EvCoKW}.

\end{thebibliography}
\end{document}